\documentstyle[preprint,epsfig,aps]{revtex}

\epsfverbosetrue
\draft

\begin{document}
\title{Structure of Si$_{12}$ Cluster}
\author{Atul Bahel and Mushti V. Ramakrishna}
\address{The Department of Chemistry, New York University,
New York, NY 10003-6621.}
\date{\prb in press \today.  Manuscript code BYJ506.}
\maketitle

\begin{abstract}

Tight-binding molecular dynamic simulations have revealed that Si$_{12}$
is an icosahedron with all atoms on the surface of an approximately 5
\AA~ diameter sphere.  This is the most spherical cage structure for
silicon clusters in the 2-13 atom size range.

\end{abstract}
\vspace{0.1 in}

The chemical reactions of silicon clusters with various reagents have
puzzled both experimentalists and theoreticians
\cite{Elkind:87,Jarrold:89,Phillips:88,Jelski:88,Kaxiras:89,Bolding:90,Patterson:90,Swift:91,Roth:94,RK:94}.
Experimentally, Si$_N$ clusters containing $N$ = 2-50 atoms have
exhibited strong oscillations in reactivities as a function of cluster
size \cite{Elkind:87,Jarrold:89}.  Theoreticians have attempted to
explain this phenomenon either through the determination of cluster
structures or through educated speculations
\cite{Phillips:88,Jelski:88,Kaxiras:89,Bolding:90,Patterson:90,Swift:91,Roth:94,RK:94}.
Despite these efforts a consistent view of the cluster structures and
reactivities has not emerged.

Recently, we started to investigate this problem by calculating the
structures of small silicon clusters using the tight-binding molecular
dynamics (TB-MD) method developed by Menon and Subbaswamy
\cite{Menon:91,Menon:93-1,Menon:93-2,Ordejon:94}.  This all valence
electron method is equivalent to the extended H\"{u}ckel method well
known in theoretical chemistry \cite{Lowe:78}.  This method has yielded
structures in excellent agreement with {\em ab initio} electronic
structure calculations for both carbon and silicon clusters
\cite{Menon:91,Menon:93-1,Menon:93-2,Ordejon:94,Tomanek:86}.  The
particular Hamiltonian we employ consists of Harrison's universal
tight-binding parameters \cite{Harrison:80}, supplemented with two
additional parameters that are fit to the bond length and vibrational
frequency of Si$_{2}$ \cite{Ordejon:94}.  Full details of this method are
described elsewhere \cite{Ordejon:94}.  Using this method in
combination with classical molecular dynamics \cite{Allen:87} and a
slow annealing schedule we optimized the structures of silicon clusters
up to $N = 13$.  The cluster structures thus derived for sizes up to $N
= 11$ are in complete agreement with the {\em ab initio} calculations
of Raghavachari and co-workers \cite{Krishnan:85,Rohlfing:90}.  This is
remarkable considering that the parameters in this tight-binding
Hamiltonian are not fit to any of the cluster structures
\cite{Ordejon:94}.  This test verifies the reliability and
predictability of our TB-MD method for the study of the structures of
larger silicon clusters.

We now focus our attention on $N$ = 11-13 clusters because experiments
with ethylene show the most dramatic variations in reactivities in this
size range, with Si$_{12}$ being most reactive and Si$_{13}$ being least
reactive, even though all these clusters are equally abundant
\cite{Jarrold:89}.  We found that Si$_{11}$ is a tetragonal antiprism
with one cap at the bottom and two caps at the top.  Rohlfing and
Raghavachari found this structure to be one of two possible candidates
for the ground state structure of Si$_{11}$ \cite{Rohlfing:90}.  The
structure of Si$_{13}$ may be described either as a 1-5-6-1 layered
structure or as a capped trigonal antiprism, similar to that obtained
by R\"{o}thlisberger and co-workers \cite{Roth:92}, but unlike the atom
centered icosahedron proposed by Chelikowsky \cite{Chelikowsky:89}.
Even though these two structures are interesting in themselves, neither
of them are unique in their geometry.

The TB-MD simulated annealing calculations also revealed that the
lowest energy structure of Si$_{12}$ is an icosahedron with a diameter of
$\approx$ 5 \AA.  Distortion from the ideal icosahedral geometry is
small.  Such a highly spherical cage structure for this cluster had not
been anticipated before.  The cohesive energy of this structure is
-3.75 eV/atom, its band gap is 1.8 eV, and its average coordination
number is 5.0.  Small distortions away from the ideal structure do not
significantly change the cohesive energy.  The ground state is a
singlet in all cases, but the perfect icosahedron has a smaller band
gap (0.26 eV) compared to its distorted form (1.8 eV).  We have also
investigated several other geometries for this cluster and found that
they are about 0.4-0.7 eV higher in energy than the icosahedron.
Hence, several structural isomers of Si$_{12}$ are possible, with the
icosahedron being the ground state structure.  The local coordination
of atoms in the icosahedral form of Si$_{12}$ is similar to that of the
axial atoms in the pentagonal bipyramidal ground state of Si$_7$.

Our calculations also show that the cohesive energies of Si$_{12}$ and
Si$_{13}$ are nearly the same, indicative that reactivity differences
between these two clusters are not related to their structural
stabilities.  However, Si$_{11}$ and Si$_{13}$ are either prolate or
asymmetric tops, whereas Si$_{12}$ is a spherical top.  This finding is
consistent with the observation of Jarrold and co-workers that
spherical clusters are at least an order of magnitude more reactive
than the non-spherical clusters \cite{Jarrold:92}.  This is probably
because the spherical surface provides the maximum number of nearest
neighbor pairs of silicon atoms that are needed to bind the ethylene
molecule.

Since Si$_{12}$ is highly reactive it has not aroused as much theoretical
interest as the inert clusters.  Consequently, until now the structure
of Si$_{12}$ has not been investigated through accurate calculations.
The few calculations \cite{Barojas:86,Feuston:87} that do exist on
Si$_{12}$ obtained structures consisting of four- and five-membered rings
that are very different from ours.  However, these calculations
employed the Stillinger-Weber \cite{Stillinger:85} or other empirical
potentials that almost always yielded incorrect cluster structures,
because these potentials were derived by fitting to the properties of
bulk Si.  Our tight-binding method does not suffer from this limitation
because all the valence electrons are explicitly included in the
calculation of the electronic energy
\cite{Menon:91,Menon:93-1,Menon:93-2,Ordejon:94,Tomanek:86}.

The icosahedral geometry with an atom at the center gives the closest
packing of thirteen hard spheres.  The atoms interacting through
two-body pair potentials typically form clusters with this structural
pattern.  For example, Ar$_{13}$ is an atom centered icosahedron
\cite{Liu:94}.  However, Ar$_{12}$ is not an icosahedron with an empty
cage \cite{Liu:94}.  The metal clusters also prefer compact structures
and do not allow empty cages \cite{Roth:92,Roth:91}.  Thus the
icosahedral cage structure is unique only for Si$_{12}$ among elemental
clusters.  The compound clusters boranes and carboranes are the only
other clusters that form icosahedral cage structures.

If two-body interaction is dominant in our tight-binding Hamiltonian
then we would expect to obtain an atom centered icosahedron for
Si$_{13}$.  However, our calculations show such a structure to be about
2 eV higher in energy compared to the lowest energy structure.
Furthermore, the structures of other Si$_N$ clusters in the 3-11 size
range are also not similar to those of Ar$_N$ clusters.  These two
observations indicate that our tight-binding Hamiltonian includes the
necessary many-body interactions required to describe the silicon
cluster structures correctly.

In summary, our tight-binding molecular dynamic simulations predict
icosahedral cage structure for Si$_{12}$.  Such a highly symmetric
non-crystalline cage structure has not been discovered previously for
any other twelve atom elemental cluster.

This research is supported by the New York University and the Donors of
The Petroleum Research Fund (ACS-PRF \# 26488-G), administered by the
American Chemical Society.

\begin{figure}
\caption{The lowest energy structure of the Si$_{12}$ cluster.  It is a
bi-capped pentagonal antiprism (BPA), called an icosahedron.  This
geometry is also describable as a 1-5-5-1 layered structure.  Experiments
have shown this cluster to be highly reactive [2].
}

\end{figure}


\begin{thebibliography}{1}

\bibitem[1]{Elkind:87} J. L. Elkind, J. M. Alford, F. D. Weiss, R. T.
Laaksonen, and R. E. Smalley, J. Chem. Phys. {\bf 87}, 2397 (1987);
S.  Maruyama, L. R. Anderson, and R. E. Smalley,
J. Chem. Phys. {\bf 93}, 5349 (1990);
J. M. Alford, R. T. Laaksonen, and R. E.  Smalley,
J.  Chem. Phys. {\bf 94}, 2618 (1991);
L. R. Anderson, S. Maruyama, and R. E. Smalley,
Chem. Phys. Lett. {\bf 176}, 348 (1991).

\bibitem[2]{Jarrold:89}
M. F. Jarrold, J. E. Bower, and K. M. Creegan, \jcp {\bf 90}, 3615 (1989);
K. M. Creegan and M. F. Jarrold, J. Am. Chem. Soc. {\bf 112}, 3768 (1990);
M. F. Jarrold, U. Ray, and K. M. Creegan, \jcp {\bf 93}, 224 (1990);
U. Ray and M. F. Jarrold, \jcp {\bf 94}, 2631 (1991).

\bibitem[3]{Phillips:88} J. C. Phillips, J. Chem. Phys. {\bf 88}, 2090
(1988).

\bibitem[4]{Jelski:88} D. A. Jelski, Z. C. Wu, and T. F. George,
Chem. Phys. Lett. {\bf 150}, 447 (1988).

\bibitem[5]{Kaxiras:89} E. Kaxiras, Chem. Phys. Lett. {\bf 163}, 323
(1989); Phys. Rev. Lett. {\bf 64}, 551 (1990).

\bibitem[6]{Bolding:90} B. C. Bolding and H. C. Andersen,
Phys. Rev. B 41, 10568 (1990).

\bibitem[7]{Patterson:90} C. H. Patterson and R. P. Messmer,
Phys. Rev.  B {\bf 42}, 7530 (1990).

\bibitem[8]{Swift:91} B. L. Swift, D. A. Jelski, D. S. Higgs, T. T. Rantala,
and T. F. George, Phys. Rev. Lett. {\bf 66}, 2686 (1991);
D. A. Jelski, B. L. Swift, T. T. Rantala, X. Xia, T. F. George,
J. Chem. Phys. {\bf 95}, 8552 (1991).

\bibitem[9]{Roth:94} U. R\"{o}thlisberger, W. Andreoni, M. Parrinello,
\prl {\bf 72}, 665 (1994).

\bibitem[10]{RK:94} M. V. Ramakrishna and J. Pan, \jcp {\bf 101}, 8108 (1994);
J. Pan and M. V. Ramakrishna, \prb {\bf 50}, 15431 (1994).

\bibitem[11]{Menon:91} M. Menon and K. R. Subbaswamy,
\prl {\bf 67}, 3487 (1991); Int. J. Mod. Phys. B {\bf 6}, 3839 (1992).

\bibitem[12]{Menon:93-1} M. Menon, K. R. Subbaswamy, and M. Sawtarie,
\prb {\bf 48}, 8398 (1993);
\prb {\bf 49}, 13966 (1994).

\bibitem[13]{Menon:93-2} M. Menon and K. R. Subbaswamy,
\prb {\bf 47}, 12754 (1993);
Chem. Phys. Lett. {\bf 219}, 219 (1994).

\bibitem[14]{Ordejon:94} P. Ordej\'on, D. Lebedenko, and M. Menon, \prb
{\bf 50}, 5645 (1994).  We use the first set of parameters with $a =
0.08$ and $b = -1.4$ described in this paper.

\bibitem[15]{Lowe:78} J. P. Lowe, {\em Quantum Chemistry},
(Academic Press, New York, 1978).

\bibitem[16]{Tomanek:86} D. Toma\'nek and M. Schl\"{u}ter,
\prl {\bf 56}, 1055 (1986); \prb {\bf 36}, 1208 (1987);
\prl {\bf 67}, 2331 (1991).

\bibitem[17]{Harrison:80} W. A. Harrison, {\em Electronic Structure
and the Properties of Solids}, (Freeman, San Francisco, 1980).

\bibitem[18]{Allen:87} M. P. Allen and D. J. Tildesley,
{\bf Computer Simulations of Liquids} (Oxford University Press, New York,
1987).

\bibitem[19]{Krishnan:85} K. Raghavachari and V. Logovinsky,
Phys. Rev.  Lett. {\bf 55}, 2853 (1985);
K. Raghavachari, J. Chem. Phys. {\bf 83}, 3520 (1985);
J. Chem. Phys. {\bf 84}, 5672 (1986);
K. Raghavachari and C. M. Rohlfing, Chem. Phys. Lett. {\bf 143}, 428 (1988);
J. Chem. Phys. {\bf 89}, 2219 (1988).

\bibitem[20]{Rohlfing:90}
C. M. Rohlfing and K. Raghavachari, Chem. Phys. Lett. {\bf 167}, 559 (1990).

\bibitem[21]{Roth:92} U. R\"{o}thlisberger, W. Andreoni, and P. Giannozzi,
\jcp {\bf 96}, 1248 (1992).

\bibitem[22]{Chelikowsky:89} J. R. Chelikowsky, J. C. Phillips, M. Kamal,
and M. Strauss, Phys. Rev. Lett. {\bf 62}, 292 (1989);
J. R. Chelikowsky, K. M. Glassford, and J. C. Phillips,
Phys. Rev. B {\bf 44}, 1538 (1991).

\bibitem[23]{Jarrold:92} M. F. Jarrold and J. E. Bower, \jcp {\bf 96},
9180 (1992).

\bibitem[24]{Barojas:86} E. Blaisten-Barojas and D. Levesque,
\prb {\bf 34}, 3910 (1986).

\bibitem[25]{Feuston:87} B. P. Feuston, R. K. Kalia, and P. Vashishta,
\prb {\bf 35}, 6222 (1987); \prb {\bf 37}, 6297 (1988).

\bibitem[26]{Stillinger:85} F. Stillinger and T. A. Weber,
\prb {\bf 31}, 5262 (1985).

\bibitem[27]{Liu:94} S. Liu, Z. Ba\v{c}i\'c, J. W. Moskowitz, and
K. E. Schmidt, \jcp {\bf 100}, 7166 (1994).

\bibitem[28]{Roth:91} U. R\"{o}thlisberger and W. Andreoni,
\jcp {\bf 94}, 8129 (1991).

\end{thebibliography}
\end{document}